\documentclass[aps,pre,preprint,groupedaddress]{revtex4}

\usepackage{graphicx}
\usepackage{latexsym}
\usepackage{amsmath,amssymb}

\begin{document}

%\preprint{}

\title{
Eliminated corrections to scaling around a renormalization-group fixed point:
Transfer-matrix simulation of an extended $d=3$ Ising model
}

\author{Yoshihiro Nishiyama}
%\thanks{}
\affiliation{Department of Physics, Faculty of Science,
Okayama University, Okayama 700-8530, Japan}

\date{\today}

\begin{abstract}
Extending the parameter space
of the three-dimensional ($d=3$) Ising model,
we search for a regime of eliminated corrections to finite-size scaling.
For that purpose, we consider a real-space renormalization group (RSRG)
with respect to
a couple of clusters simulated with the transfer-matrix (TM) method.
Imposing a criterion of ``scale invariance,''
we determine a location of the non-trivial RSRG fixed point.
Subsequent large-scale TM simulation around the fixed point
reveals eliminated corrections to finite-size scaling.
As anticipated, such an elimination of corrections admits
systematic finite-size-scaling analysis.
We obtained the estimates for the critical indices as
$\nu=0.6245(28)$ and $y_h=2.4709(73)$.
As demonstrated, with the aid of the preliminary RSRG survey,
the transfer-matrix simulation provides rather reliable information
on criticality even for $d=3$,
where the tractable system size is restricted severely.
\end{abstract}

\pacs{
64.60.Ak % Renormalization-group, fractal, and percolation studies of phase transitions 
         % (see also 61.43.Hv Fractals; macroscopic aggregates)
5.10.-a % Computational methods in statistical physics and 
        % nonlinear dynamics (see also
05.70.Jk % Critical point phenomena
75.40.Mg % Numerical simulation studies
}
%\keywords{}

\maketitle

\section{\label{section1}Introduction}

The transfer-matrix method has an advantage over the Monte Carlo method
in that it provides information
free from the statistical (sampling) error and the problem of slow relaxation
to thermal equilibrium.
On one hand,
the tractable system size with the transfer-matrix method
is severely limited, because the transfer-matrix size increases exponentially
as the system size $N$ enlarges;
here, $N$ denotes the number of spins constituting a unit of the transfer-matrix slice.
Such a limitation becomes even more serious in large dimensions ($d \ge 3$).
Actually, for large $d$, the system size $N(=L^{d-1})$ increases rapidly 
as the linear dimension $L$ enlarges, and it soon exceeds
the limit of available computer resources.
Because of this difficulty, the usage of the transfer-matrix method
has been restricted mainly within $d=2$.

In this paper,
we report an attempt to eliminate the finite-size corrections
of the $d=3$ Ising model by tuning the interactions parameters.
As anticipated, such an elimination of corrections admits
systematic finite-size-scaling analysis of the numerical data
with restricted system sizes.
To be specific, we consider the 
$d=3$ Ising ferromagnet with the extended interactions,
\begin{equation}  
\label{Hamiltonian}
H=-J_{NN} \sum_{\langle i,j \rangle} S_i S_j 
   -J_{NNN} \sum_{\langle \langle i, j \rangle\rangle} S_i S_j 
   -J_{\square} \sum_{[i,j,k,l]} S_i S_j S_k S_l
 ,
\end{equation}
where the Ising spins $S_i = \pm 1$ are placed at the cubic-lattice points 
specified by the index $i$;
The summations 
$\sum_{\langle i,j\rangle}$,
$\sum_{\langle \langle i,j \rangle\rangle}$, and
$\sum_{[i,j,k,l]}$ run over all nearest-neighbor pairs, 
next-nearest-neighbor (plaquette diagonal) spins, and 
round-a-plaquette spins, respectively.
Within the extended parameter space $(J_{NN},J_{NNN},J_\square)$, we search for a regime of eliminated
corrections to scaling.
For that purpose, we consider a 
real-space renormalization group
for a couple of clusters, whose
thermodynamics is simulated with the transfer-matrix method; see Fig. \ref{figure1}.
We then determine a location of the renormalization-group fixed point.
Following this preliminary renormalization-group survey, we 
perform extensive transfer-matrix simulation around this fixed point.
Thereby, we show that the corrections-to-scaling behavior is improved around the fixed point.
Here, we utilized an improved version of the transfer-matrix method 
\cite{Novotny90,Novotny92,Novotny93,Novotny91,Nishiyama04,Nishiyama05,Nishiyama06},
and succeeded in treating 
a variety of system sizes $N=5,6,\dots,15$;
note that conventionally, the tractable system sizes are restricted to $N=4,9,16,\dots$.
Apparently, such an extension of available system sizes provides 
valuable information on criticality.

In fairness, it has to be mentioned that our research owes its basic idea to 
the following pioneering studies:
First, an attempt to eliminate the finite-size corrections was reported in
Ref. \cite{Blote96}, where the authors investigate
the $d=3$ Ising model with the (finely-tuned) second and third
neighbor interactions;
see also the studies 
\cite{Ballesteros98,Hasenbusch99,Hasenbusch00}
in the lattice-field-theory context.
Their consideration could be viewed as an interesting application of 
the Monte Carlo renormalization group \cite{Swendsen82} to
exploit the virtue of the fixed point.
(The Monte Carlo renormalization group provides an explicit realization of the renormalization-group
idea
in the real space.)
The aim of this paper is to
develop an alternative approach to the elimination of corrections
via the transfer-matrix method, 
and make the best use of its merits and characteristics.
In fact,
the four-spin interaction, appearing in our Hamiltonian (\ref{Hamiltonian}),
 is readily tractable with the transfer-matrix method,
whereas it
can make a conflict with the Monte Carlo simulation;
in fact,
the four-spin interaction disables the use of cluster update.
(Probably, as for the Monte Carlo simulation, 
it might be more rewarding to enlarge the system size rather than 
incorporate extra interactions.)
Second, the extended interactions appearing in our Hamiltonian (\ref{Hamiltonian})
are taken from the proposal by Ma \cite{Ma76},
who investigated the $d=2$ Ising model and its renormalization-group flow.
We consider that his renormalization-group scheme for $d=2$ is still 
of use to our $d=3$ case as well.
Actually, in our transfer-matrix treatment, the system size along the transfer-matrix direction
is infinite, and the remaining $d=2$ fluctuations are responsible for the finite-size corrections.
We demonstrate that Ma's scheme leads to satisfactory elimination of
finite-size corrections in $d=3$.

The rest of this paper is organized as follows.
In Sec. \ref{section2}, we explain the real-space decimation (renormalization group) for the
$d=3$ Ising model (\ref{Hamiltonian}), and search for its fixed point.
In Sec. \ref{section3}, 
we perform extensive transfer-matrix simulation around this fixed point.
Here, we utilized an improved transfer-matrix method,
 which is explicated in Appendix \ref{appendix}.
The last section is devoted to summary and discussions.

\section{\label{section2}
Search for a scale-invariant point: A regime of eliminated irrelevant interactions}

In this section, we search for a point of eliminated finite-size corrections
of the extended Ising model, Eq. (\ref{Hamiltonian}).
For that purpose, we set up a real-space renormalization group,
and look for the scale-invariant (fixed) point;
the result is given by Eq. (\ref{fixed_point}).

\begin{figure}
\includegraphics[width=100mm]{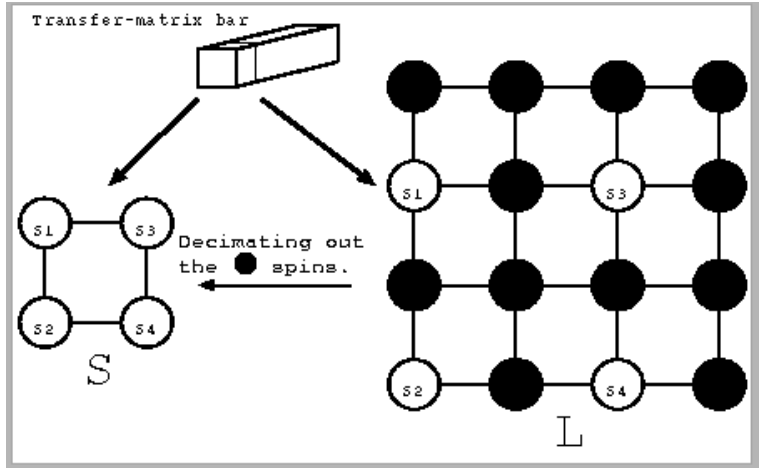}%
\caption{\label{figure1}
A schematic drawing of our real-space renormalization group (decimation)
for the $d=3$ Ising model with the extended interactions, Eq. (\ref{Hamiltonian}).
As indicated, the thermodynamics is simulated with the transfer-matrix method.
Imposing a criterion of scale invariance, Eq. (\ref{scale_invariance}), 
we determined the renormalization-group fixed point, Eq. (\ref{fixed_point}),
numerically; see text for details.
}
\end{figure}

To begin with,
we set up the real-space renormalization group.
We consider a couple of rectangular clusters with the sizes $2\times 2$ and $4 \times 4$;
see Fig. \ref{figure1}.
These clusters are labeled by the symbols $S$ and $L$, respectively.
(Because we utilize the transfer-matrix method, 
the system sizes perpendicular to these rectangles are both infinite.)
Decimating out the spin variables indicated by the symbol $\bullet$ of the $L$ cluster,
we obtain a reduced lattice structure identical to that of the $S$ cluster.
Our concern is to find a ``scale invariance'' condition with respect to this real-space
renormalization group.

Before going into the explicit formulation of the renormalization group
(fixed-point analysis),
we explain briefly how we simulated the thermodynamics of these clusters.
As mentioned above, we employ the transfer-matrix method.
The transfer-matrix elements for the $S$ cluster are given by the formula,
\begin{equation}
T_{\{T_i\},\{S_i\}} =
   (W_{S_1 S_2}^{S_3 S_4})^{1+3b}
 (
    W_{S_1 S_2}^{T_1 T_2}
    W_{S_2 S_4}^{T_2 T_4}
    W_{S_4 S_3}^{T_4 T_3}
    W_{S_3 S_1}^{t_3 T_1}
  )^{1+b}   ,
\end{equation}
where the component $W_{S_1 S_2}^{S_3 S_4}$ denotes the local Boltzmann weight for a 
plaquette, Eq. (\ref{local_Boltzmann_weight}),
and the spin variables $\{ S_i \}$ and $\{ T_i \}$ 
($i=1\sim4$) denote the spin configurations for both sides
of the transfer-matrix slice.
The component $(\cdots)^{1+3b}$ originates in the plaquette interactions perpendicular to the
transfer-matrix direction, whereas
the remaining part $(\cdots)^{1+b}$ comes from the longitudinal ones.
The parameter $b$ controls the boundary-interaction strength.
Note that irrespective of $b$, the periodic-boundary condition is maintained;
namely, all $2 \times 2$ spins remain equivalent as $b$ varies.
Such a redundancy is intrinsic to the $L=2$ system.
Here, we consider this redundant parameter as a freely tunable one.
(For example, a naive implementation of the periodic-boundary condition for a pair of spins
may result in such an interaction as 
$H=-J S_1 S_2 -J S_2 S_1=-2 J S_1 S_2$.
Apparently, such a duplicated interaction is problematic.
Possibly, the interaction $-(1+b)J S_1 S_2$ with a certain moderate parameter $b$ 
should be a favorable one.
Significant point is that the periodic boundary condition is maintained
with $b$ varied.)
We found that the choice $b=0.4$ is reasonable because of the reasons mentioned afterward.

Similarly to the above, we constructed the transfer matrix for the $L$ cluster as,
\begin{equation}
T_{\{T_{ij}\},\{S_{ij}\}}=
\prod_{1 \le i,j \le 4} 
(W_{S_{ij}S_{i+1,j}}^{S_{i,j+1}S_{i+1,j+1}}
W_{S_{ij}S_{i+1,j}}^{T_{ij}T_{i+1,j}}
W_{S_{ij}S_{i,j+1}}^{T_{ij}T_{i,j+1}})
,
\end{equation}
with the $4 \times 4$ spin configurations $\{ S_{ij} \}$ and $\{ T_{ij} \}$ under the periodic
boundary condition.
In this case ($L=4$), we have no ambiguity as to the boundary interaction.

Based on the above-mentioned simulation scheme, we calculate 
the location of the renormalization-group fixed point.
We impose the following ``scale-invariance'' conditions,
\begin{eqnarray}
\label{scale_invariance}
\langle S_1 S_2 \rangle_S &=& \langle S_1 S_2 \rangle_L \\
\langle S_1 S_4 \rangle_S &=& \langle S_1 S_4 \rangle_L \\
\langle S_1 S_2 S_3 S_4\rangle_S &=& \langle S_1 S_2 S_3 S_4\rangle_L .
\end{eqnarray}
Here, the symbol $\langle \dots \rangle_{S(L)}$ denotes the thermal average for the $S$ ($L$) cluster,
and the arrangement of spin variables $\{ S_i \}$ is shown in Fig. \ref{figure1}.
We solve the solution of the above equations numerically,
and found that a non-trivial solution does exist at,
\begin{equation}
\label{fixed_point}
(\tilde{J}_{NN},\tilde{J}_{NNN},\tilde{J}_\square)=
(0.108982866643  5 % 646   
,
0.04 45777727956 %% 07515E-002
,
-0.0065117950492 % 092920E-003
)  .
\end{equation}
The last digits may be uncertain due to the numerical round-off errors.
The result is to be compared with that of the
preceding Monte Carlo study
$(\tilde{J}_{NN},\tilde{J}_{NNN},\tilde{J}_{3rd})=
(0.1109,0.03308,0.01402)$ \cite{Blote96},
where the authors 
incorporated 
the third-neighbor interaction $\tilde{J}_{3rd}$ and omitted $\tilde{J}_\square$ instead.

Let us mention a few comments.
First, in the next section, we confirm that the fixed point is 
indeed a good approximant to
the phase-transition point.
This fact indicates that the above renormalization-group analysis is 
sensible.
Moreover,
we calculated
the fixed point $\tilde{J}_{NN}=0.2243904423106$ for the conventional Ising model 
$(J_{NNN},J_\square)=(0,0)$.
We again see that this transition point is in agreement with a critical point
$J_{NN}^*=0.22165455(3)$ determined with the Monte Carlo method \cite{Deng03}.
(Hence, the choice of the boundary-interaction strength $b=0.4$
is justified.)
Second, we stress that the above renormalization group is not
intended to obtain (quantitatively reliable) critical point nor the critical indices.
{\it The aim of the above analysis is to truncate out the irrelevant interactions.}
The detailed analysis on criticality is made with the subsequent
finite-size-scaling analysis.
In other words, our numerical approach consists of two steps,
and the remaining step is considered in the next section.

\section{\label{section3}
Finite-size scaling analysis of the critical exponents $\nu$ and $y_h$}

In the preceding Sec. \ref{section2}, we determined the position of 
the renormalization-group fixed point, Eq. (\ref{fixed_point}).
In this section, around the fixed point, we survey the criticality of 
the {\it temperature-driven} phase transition.
Namely, hereafter,
we dwell on the one-parameter subspace,
\begin{equation}
\label{parameter_space}
(J_{NN},J_{NNN},J_\square)=J_{NN}  
 \left(
              1,\frac{\tilde{J}_{NNN}}{\tilde{J}_{NN}},
                \frac{\tilde{J}_\square}{\tilde{J}_{NN}}
   \right),
\end{equation}
which contains
the renormalization-group fixed point at $J_{NN}=\tilde{J}_{NN}$.
We anticipate that corrections to scaling (influence of irrelevant operators) 
should be suppressed
in this parameter space.

Throughout this section, we employ an improved version of the transfer-matrix method
(Novotny's method) \cite{Novotny90}.
(To avoid confusion, we stress that 
in the above section, we used the conventional transfer-matrix method.)
A benefit of Novotny's method is that
we are able to treat arbitrary (integral) number of spins, $N=5,6,\dots,15$,
constituting a unit of the transfer-matrix slice;
note that conventionally, the number of spins is restricted to $N=4,9,16,\dots$.
We explicate this simulation algorithm in Appendix \ref{appendix}.

\subsection{\label{section3_1}
Eliminated corrections to scaling}

\begin{figure}
\includegraphics[width=100mm]{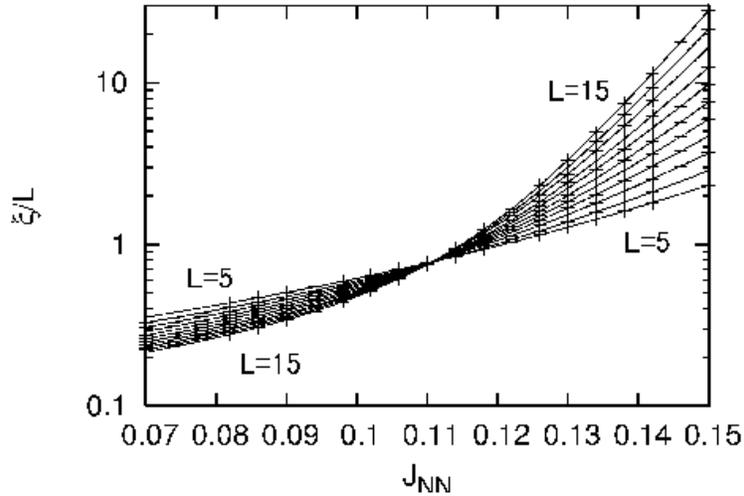}%
\caption{\label{figure2}
Scaled correlation length $\xi/L$ is plotted for the nearest-neighbor interaction $J_{NN}$
and $N=5,6,\dots,15$ ($N=L^2$);
note that we survey the parameter space (\ref{parameter_space}) including
the renormalization-group fixed point (\ref{fixed_point}).
We observe a clear indication of criticality at $J_{NN} \sim 0.11$.
Apparently, the finite-size-scaling behavior is improved 
as compared to that of 
Fig. \ref{figure4} for the conventional Ising model.
}
\end{figure}

In Fig. \ref{figure2}, we plotted the scaled correlation length $\xi/L$ for 
$J_{NN}$ and a variety of system
sizes $N=5,6,\dots,15$.
We evaluated the correlation length $\xi$ with use of the formula
$\xi=1/\ln(\lambda_1/\lambda_2)$
with the dominant (sub-dominant) eigenvalue $\lambda_1$ ($\lambda_2$)
of the transfer matrix.
As explained in Appendix \ref{appendix}, the linear dimension $L$ is simply given by,
\begin{equation}
L=\sqrt{N}   ,
\end{equation}
with the number of spins $N$; see Fig. \ref{figure8}.

From Fig. \ref{figure2}, we see a clear indication of criticality at $J_{NN} \approx 0.11$;
note that the intersection point of the curves
indicates a critical point.
Afterward, we compare this result to that of the conventional Ising model 
$(J_{NNN},J_\square)=(0,0)$
to elucidate an improvement of the scaling behavior.
Here, we want to draw reader's attention to the point that we treated 
various system sizes $N=5,6,\dots,15$ with the aid of the Novotny method.
Actually, in Fig. \ref{figure2}, we notice that a variety of system sizes
are available.
Apparently, such an extension of available system sizes 
is significant in the subsequent detailed finite-size-scaling analyses.

\begin{figure}
\includegraphics[width=100mm]{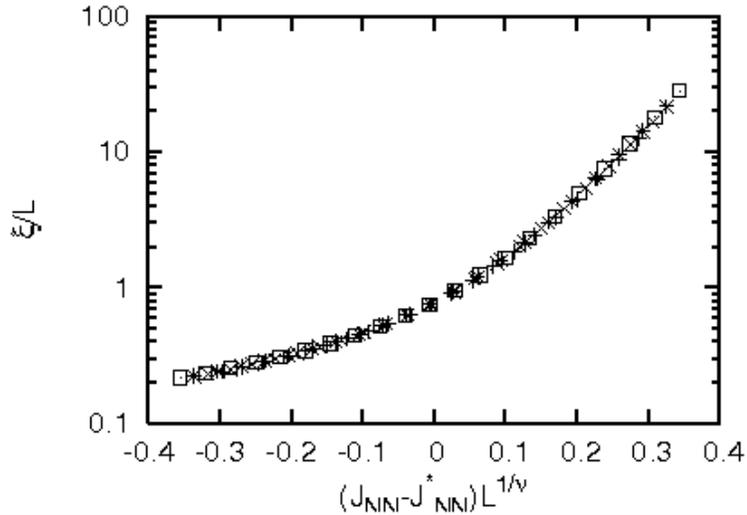}%
\caption{\label{figure3}
The scaling plot for the correlation length, $(J_{NN}-J_{NN}^*)L^{1/\nu}$-$\xi/L$,
is shown for the system sizes ($+$) $N=12$, ($\times$) 13,
($*$) 14, and ($\square$) 15; note the relation $N=L^2$.
Here, we accepted the scaling parameters, $J_{NN}^*=0.11059$ and $\nu=0.6245$, determined in Figs.
\ref{figure5} and \ref{figure6}, respectively.
We again confirm that corrections to scaling are suppressed significantly.}
\end{figure}

In Fig. \ref{figure3},
we presented the scaling plot
 $(J_{NN}-J_{NN}^*)L^{1/\nu}$-$\xi /L$ for $12 \le N \le 15$.
with the scaling parameters $J_{NN}^*=0.11059$ and $\nu=0.6245$ determined 
in Figs. \ref{figure5} and \ref{figure6}, respectively.
We see that the data collapse into a scaling function satisfactorily;
actually, we can hardly observe corrections to the finite-size scaling.

\begin{figure}
\includegraphics[width=100mm]{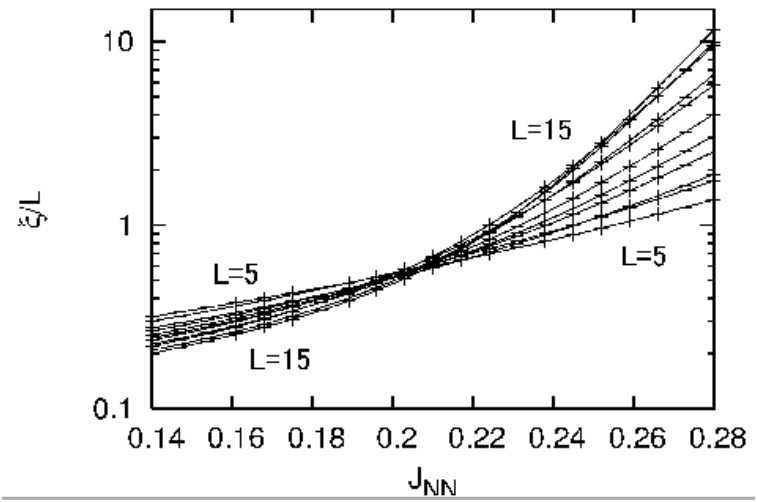}%
\caption{\label{figure4}
Tentatively, we turned off the 
extended interactions ($J_{NNN}=0$ and $J_\square=0$),
and calculated the
scaled correlation length $\xi/L$ for various $J_{NN}$
and $N=5,6,\dots,15$.
We notice that the data are scattered as compared to those in Fig. \ref{figure2}.
}
\end{figure}

In the above, we presented an evidence that the corrections-to-scaling behavior is
improved in the parameter space, Eq. (\ref{parameter_space}).
Lastly, as a comparison,
we provide the data for the conventional Ising model;
namely, we set $(J_{NNN},J_\square)=(0,0)$ tentatively.
In Fig. \ref{figure4},
we plotted the scaled correlation length $\xi/L$ for various $J_{NN}$.
Apparently, the data suffer from insystematic finite-size corrections.
The data scatter obscures the position of
critical point.
(Nevertheless, we should mention that the data
imply $J_{NN}^* \sim 0.2$, which does not
contradict a recent Monte Carlo result $J_{NN}^*=0.22165455(3)$ \cite{Deng03}.)

\subsection{Phase-transition point $J_{NN}^*$}

\begin{figure}
\includegraphics[width=100mm]{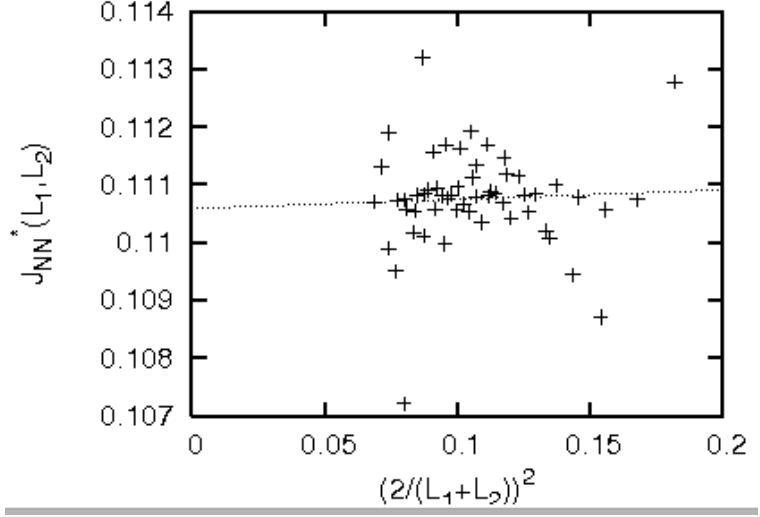}%
\caption{\label{figure5}
The approximate critical interaction $J_{NN}^*(L_1,L_2)$
is plotted for $(2/(L_1+L_2))^2$
with $5 \le N_1 <  N_2 \le 15$ ($L_{1,2}=\sqrt{N_{1,2}}$).
The least-squares fit to these data yields $J_{NN}^*=0.11059(52)$
in the thermodynamic limit $L \to \infty$.
}
\end{figure}

In the above, we obtain a rough estimate for the phase-transition point $J_{NN}^*$.
In this section, we determine the transition point more precisely.
In Fig. \ref{figure5}, we plotted the approximate transition point
$J_{NN}^*(L_1,L_2)$ for $(2/(L_1+L_2))^2$.
Here, the approximate transition point denotes the intersection point
of the curves $\xi/L$ (Fig. \ref{figure2}) for a pair of 
system sizes $(L_1,L_2)$ ($5 \le N_1 < N_2 \le 15$).
Namely, the following equation,
\begin{equation}
\label{approximate_critical_point}
 \left. L_1 \xi(L_1) \right|_{J_{NN}=J_{NN}^*(L_1,L_2)} = 
 \left. L_2 \xi(L_2) \right|_{J_{NN}=J_{NN}^*(L_1,L_2)}  ,
\end{equation}
holds.
In Fig. \ref{figure5}, we notice that
the data exhibit suppressed systematic finite-size deviation.
Namely, the insystematic data scatter is more conspicuous
than the systematic deviation.
The least-squares fit to these data yields the transition point,
\begin{equation}
\label{transition_point}
J_{NN}^*=0.11059(52) ,
\end{equation}
in the thermodynamic limit $L \to \infty$.

In order to check the reliability, we replaced the abscissa scale with 
$(2/(L_1 + L_2))^{\omega+1/\nu}$ \cite{Binder81},
where we used $1/\nu=1.5868(3)$ and $\omega=0.821(5)$ reported in Ref. \cite{Deng03}.
(In the next section, we make a consideration on the abscissa scale.)
Thereby, we arrive at $J_{NN}^*=0.11062(43)$,
which is consistent with the above result.
(The error margin may come from purely statistical one.) 
We confirm that the choice of the abscissa scale is not so influential.
%This fact indicates that the above extrapolated value would contain little 
%systematic (biased) error.

We notice that the transition point $J_{NN}^*$ (\ref{transition_point}) 
and the renormalization-group fixed
point $\tilde{J}_{NN}$ (\ref{fixed_point}) are in good agreement with each other.
This fact confirms that the renormalization-group analysis in Sec. \ref{section2}
is indeed sensible.
As mentioned in Sec. \ref{section2}, we do not require fine accuracy as to the convergence
of $J_{NN}^*$ and $\tilde{J}_{NN}$. 
The aim of the renormalization-group analysis
is to search for a regime of eliminated corrections rather than to obtain
the precise location of the fixed point.
Detailed analysis on criticality 
is performed in the subsequent finite-size-scaling analysis
as demonstrated in the next section.
(Actually, tuning the boundary-interaction parameter $b$ (see Sec. \ref{section2}), 
we could attain better agreement between $J_{NN}^*$ and $\tilde{J}_{NN}$.
However, such a refinement does not affect the subsequent finite-size-scaling analysis
very much.)

\subsection{Critical exponents $\nu$ and $y_h$}

In Sec. \ref{section3_1}, we presented an evidence of eliminated finite-size corrections.
Encouraged by this result, in this section,
we evaluate the critical exponents $\nu$ and $y_h$
with use of the finite-size-scaling method
(phenomenological renormalization group) \cite{Nightingale76}.

\begin{figure}
\includegraphics[width=100mm]{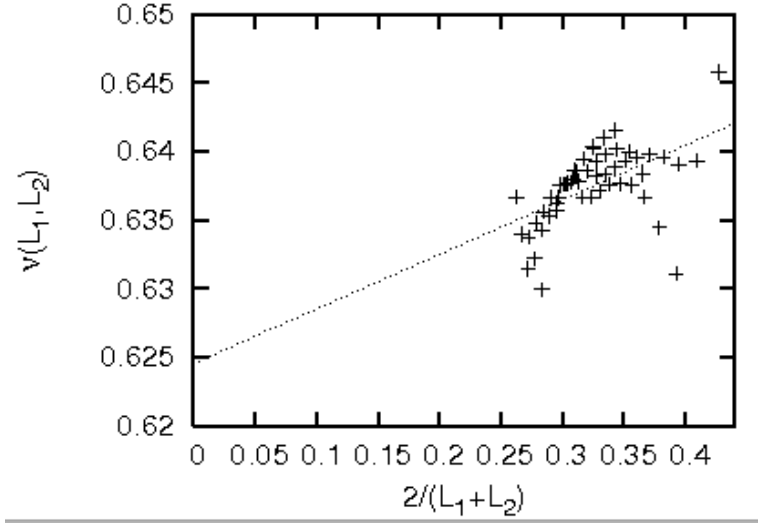}%
\caption{\label{figure6}
The approximate correlation-length critical exponent $\nu(L_1,L_2)$
is plotted for $2/(L_1+L_2)$ with $5 \le N_1 < N_2 \le 15$
($L_{1,2}=\sqrt{N_{1,2}}$).
The least-squares fit to these data yields 
$\nu=0.6245(28)$
in the thermodynamic limit $L \to \infty$.
}
\end{figure}

In Fig. \ref{figure6}, we plotted the approximate correlation-length critical exponent,
\begin{equation}
\label{exponent_nu}
\nu(L_1,L_2) = \ln(L_1   /   L_2) / 
    \left.  \ln \left(
     \frac{\partial (\xi(L_1)/L_1)}{\partial J_{NN}}   /
     \frac{\partial (\xi(L_2)/L_2)}{\partial J_{NN}}
      \right)  \right|_{J_{NN}=J_{NN}^*(L_1,L_2)}           ,
\end{equation}
for $2/(L_1+L_2)$ with $5 \le N_1 < N_2 \le 15$.
With use of
the least-squares fit to these data,
we obtain the estimate,
\begin{equation}
\nu=0.6245(28),
\end{equation}
 in the thermodynamic limit.
The data in Fig. \ref{figure6} exhibit appreciable
systematic finite-size corrections.
More specifically, 
the systematic deviation ($\sim 2 \%$) is almost comparable to 
the insystematic data scatter.
(This fact indicates that we cannot fully truncate out
the irrelevant interactions within the parameter space 
$(J_{NN},J_{NNN},J_\square)$.)
In this sense,
the above (extrapolated) value, Eq. (\ref{exponent_nu}),
may contain a systematic (biased)
error.
Afterward,
we make a few considerations on the extrapolation scheme.
(Because our work is methodology-oriented,
we supply the least-squares-fit result as it is.)

\begin{figure}
\includegraphics[width=100mm]{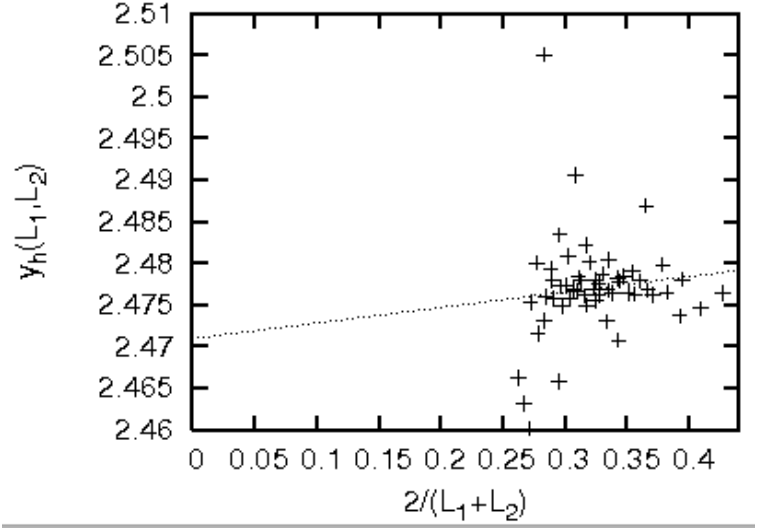}%
\caption{\label{figure7}
The approximate critical exponent $y_h(L_1,L_2)$
is plotted for $2/(L_1+L_2)$ with $5 \le N_1 < N_2 \le 15$
($L_{1,2}=\sqrt{N_{1,2}}$).
The least-squares fit to these data yields 
$y_h=2.4709(73)$
in the thermodynamic limit $L \to \infty$.
}
\end{figure}

In Fig. \ref{figure7}, we plotted the approximate exponent $y_h$,
\begin{equation}
y_h(L_1,L_2) = 
   \ln
 \left.   \left(
     \frac{\partial^2 (\xi(L_1)/L_1)}{\partial H^2}  /
     \frac{\partial^2 (\xi(L_2)/L_2)}{\partial H^2}
      \right)  \right|_{J_{NN}=J_{NN}^*(L_1,L_2)}   
       / (2 \ln(L_1/L_2) )   
                      ,
\end{equation}
for $2/(L_1+L_2)$ with $5 \le N_1 < N_2 \le 15$.
In order to incorporate the magnetic field $H$,
we added the Zeeman term, $-H \sum_i S_i$, 
to the Hamiltonian (\ref{Hamiltonian}).
Rather satisfactorily,
the data $y_h(L_1,L_2)$ exhibit suppressed systematic corrections;
the systematic deviation is almost negligible compared to the insystematic one.
The least-squares fit to these data yields the estimate,
\begin{equation}
y_h=2.4709(73),
\end{equation}
 in the thermodynamic limit.

Provided by the above estimates $\nu$ and $y_h$,
we obtain the following critical indices through the scaling relations;
\begin{eqnarray}
\alpha &=& 0.1265(84)\\
\beta  &=& 0.3304(48)\\
\gamma &=& 1.213(11)         .
\end{eqnarray}

Let us provide comparative results with an alternative extrapolation scheme.
We replaced the scale of
abscissa in Figs. \ref{figure6} and \ref{figure7} with $(2/(L_1+L_2))^\omega$;
here, we set the exponent $\omega=0.821(5)$ reported in Ref. \cite{Deng03}.
(As mentioned below, this scheme may overestimate the amount of systematic
finite-size corrections.)
Accepting this abscissa scale,
we arrive at $\nu=0.6216(34)$ and $y_h=2.4694(90)$.
These values appear to be consistent with the above ones within the error margins,
confirming that the extrapolation scheme is not so influential.

We argue the underlying physics of the abscissa scale (extrapolation scheme)
in detail.
In principle,
the exponent $\omega$ governs the dominant (systematic) finite-size corrections.
On the other hand,
in the present simulation,
we are trying to truncate out such systematic corrections.
Hence, in our data analysis, the usage of the exponent $\omega$ would be problematic.
We consider that the systematic corrections should obey the scaling law
like $L^{-\omega_{eff}}$ with a certain effective exponent $\omega_{eff}>\omega$,
at least, in the regime of $5 \le N \le 15$.
Namely, we suspect that
the abscissa scale with the exponent $\omega$ leads to 
an overestimation of systematic corrections.
Actually, a recent Monte Carlo simulation reports the 
estimates $\nu=0.63020(12)$ and $y_h=2.4816(1)$ \cite{Deng03}.
Here, we notice that
their $\nu$ indicates a non-negligible deviation,
whereas the value of
$y_h$ is in good agreement with ours. 
This fact confirms the above observation that $\nu(L_1,L_2)$
exhibits appreciable systematic corrections,
and the extrapolated value may contain a biased error.
Possibly,
the adequate exponent $\omega_{eff}$ would be even larger
than the value $\omega_{eff}=1$ utilized in Fig. \ref{figure6}.
Nevertheless, for the sake of simplicity,
we do not pursue this issue further, and 
supply the least-squares-fit result as it is.

Lastly, we mention a recent extensive
exact-diagonalization result by Hamer \cite{Hamer00},
who obtained 
$\nu=0.62854(79)$
and $y_h=2.482(10)$.
%% , $\alpha=0.1144(24)$, $\beta=0.3281(13)$, and $\gamma=1.2319(65)$.
He investigated the quantum $d=2$ transverse-field Ising model,
relying on the belief that the quantum $d=2$ Ising model should belong to
the same universality class as the $d=3$ Ising ferromagnet.
The quantum version has an advantage
such that the Hamiltonian elements are sparse (few non-zero elements),
and one is able to treat a large cluster size $6 \times 6$.
Comparing our data with his results,
we notice that they are almost comparable with each other.
Actually,
the error margin of our $y_h$ is even smaller than his result,
although we treated the $d=3$ ferromagnet directly.

\section{\label{section4}Summary and discussions}

So far, it has been considered that the transfer-matrix method would not 
be very useful to the problems in $d \ge 3$ because of its severe limitation 
as to the tractable system sizes.
In this paper, we demonstrated that 
the corrections-to-scaling behavior of the $d=3$ Ising model
(\ref{Hamiltonian}) is improved by adjusting the coupling constants
to the values of the renormalization-group fixed point, Eq. (\ref{fixed_point}).
Actually, 
corrections to scaling in Figs. \ref{figure2} and \ref{figure3} are
eliminated significantly as compared to those 
in Fig. \ref{figure4} for
the conventional Ising model.
Moreover, we succeeded in treating a variety
of system sizes $N=5,6,\dots,15$ with the aid of the Novotny method 
(Appendix \ref{appendix});
note that with the conventional approach, the available system sizes are restricted
to $N=4,9,16,\dots$.
Apparently, such an extension of available system sizes provides
valuable information on criticality.
Owing to these improvements, we analyzed the criticality
of the 
$d=3$ 
Ising model with the transfer-matrix method,
and obtained 
the critical indices $\nu=0.6245(28)$ and $y_h=2.4709(73)$.

As mentioned in Introduction, an attempt to eliminate finite-size corrections has been pursued
\cite{Blote96}
in the context of the Monte Carlo renormalization group \cite{Swendsen82}. 
We consider that an approach with the transfer-matrix method is also of use
because of the following reasons.
First, we accepted a simple renormalization-group scheme 
shown in Fig. \ref{figure1}.
As mentioned in Introduction, this scheme was introduced originally 
as for the $d=2$ Ising model 
\cite{Ma76}.
The advantage of the transfer-matrix method is that
the system size along the transfer-matrix direction is infinite,
and 
the remaining $d=2$ fluctuations are responsible for the finite-size corrections.
Hence, such a ($d=2$)-like renormalization group is still of use to achieve elimination of corrections
satisfactorily.
Second, the transfer-matrix method is capable of the four-spin interaction appearing in
our Hamiltonian (\ref{Hamiltonian}).
On the other hand,
the Monte Carlo sampling conflicts with such a multi-spin interaction,
because
the multi-spin interaction disables the use of cluster-update algorithm.
(Probably, an effort toward enlarging the system size
would be rewarding from a technical viewpoint.)

In addition to these merits, we would like to emphasize again the point 
that
the transfer-matrix approach with the Novotny method
allows us to treat a variety of system sizes $N=5,6,\dots,15$.
We consider that
Novotny's method combined with the elimination of finite-size corrections
would be promising to
resolve the (seemingly intrinsic) drawback of the transfer-matrix method in $d \ge 3$.
As a matter of fact,
the basic idea of the present scheme would be generic,
and it might have a potential applicability to a wide class of systems.
An effort toward this direction is in progress,
and it will be addressed in future study.

\begin{acknowledgments}
This work is supported by a Grant-in-Aid 
(No. 15740238) from Monbu-Kagakusho, Japan.
\end{acknowledgments}

\appendix

\section{\label{appendix}Novotny's transfer-matrix method}

We explain the details of the transfer-matrix method
utilized in Sec. \ref{section3}.
(To avoid confusion, we remind the reader that in Sec. \ref{section2}, 
we utilized the conventional
transfer-matrix method.)
Our method is based on Novotny's formalism
\cite{Novotny90,Novotny92,Novotny93,Novotny91}, which enables us
to consider an arbitrary (integral) number of spins ${}^\forall N$,
constituting a unit of the transfer-matrix slice
even for $d \ge 3$; 
note that conventionally, the number of spins is restricted to $N=4,9,16,\dots$.
We made a modification to the Novotny formalism in order  
to incorporate the plaquette-type interactions.
We already reported this method in Ref. \cite{Nishiyama04},
where we studied the multicriticality of the extended $d=3$ Ising model \cite{Savvidy94}.
In the present paper, we implemented yet further modifications such as
Eqs. (\ref{modification1})-(\ref{modification3}).
Hence, for the sake of self-consistency,
we explicate the full details of the simulation scheme.

\begin{figure}
\includegraphics[width=100mm]{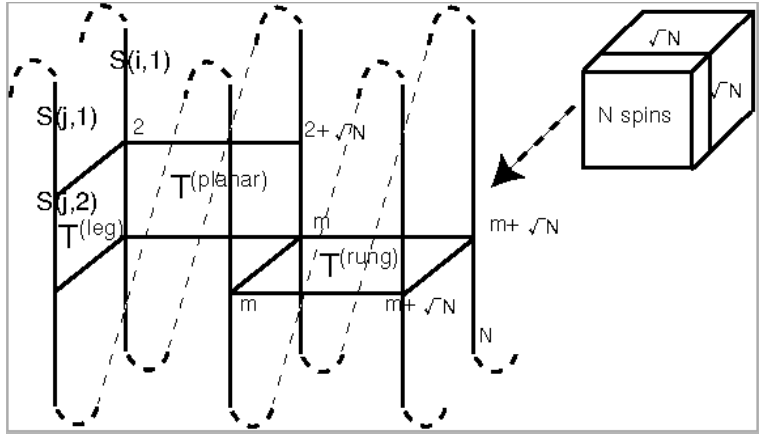}%
\caption{\label{figure8}
A schematic drawing of a unit of the transfer-matrix slice
for the $d=3$ Ising model with the extended interactions (\ref{Hamiltonian}).
The contributions from the ``leg,'' ``planar,'' and ``rung''
interactions are considered separately; see Eq. (\ref{transfer_matrix}).
Within the transfer-matrix slice, 
the arrangement of the constituent spins is one-dimensional
(coiled structure).
The dimensionality is raised up to $d=2$ by the bridges 
between the $\sqrt{N}$th-neighbor spins along the leg.
}
\end{figure}

Before going into details,
we mention the basic idea of the Novotny method.
In Fig. \ref{figure8},
We presented a schematic drawing of a unit of the transfer-matrix slice.
Note that in general, a transfer-matrix unit for a $d$-dimensional 
system
should have a $(d-1)$-dimensional structure, because it is a crosssection of
the $d$-dimensional manifold.
However, as shown in Fig. \ref{figure8},
the constituent $N$ spins form a $d=1$ (coiled)
alignment rather than $d=2$.
The dimensionality is raised {\it effectively} to $d=2$ by the
$\sqrt{N}$th-neighbor interactions among these $N$ spins;
This is the essential idea of the
Novotny method to
constitute a transfer-matrix unit with arbitrary number of spins even for $d = 3$.

In the following, we present the explicit formulas for the transfer-matrix elements.
We decompose the transfer matrix into the following three components:
\begin{equation}
\label{transfer_matrix}
T(v)=  T^{(leg)} \odot T^{(planar)}(v) \odot T^{(rung)}(v)   ,
\end{equation}
where the symbol $\odot$ denotes the Hadamard (element by element) matrix multiplication.
Note that the product of local Boltzmann weight gives rise to the global one.
The physical content of each component is shown in Fig. \ref{figure8}.

The explicit expression for the element of $T^{(leg)}$
is given by the formula,
\begin{equation}
T_{ij}^{(leg)}  =  \langle i| A | j \rangle
 = W_{S(i,1)S(i,2)}^{S(j,1)S(j,2)}
W_{S(i,2)S(i,3)}^{S(j,2)S(j,3)}
  \dots
W_{S(i,N)S(i,1)}^{S(j,N)S(j,1)}
   ,
\end{equation}
where the indices $i$ and $j$ specify the spin configurations of both sides of the transfer-matrix slice.
More specifically, the spin configuration 
$\{ S(i,1),S(i,2), \dots , S(i,N) \}$ is
arranged along the leg; see Fig. \ref{figure8}.
The factor 
$W_{S_1 S_2}^{S_3 S_4}$
denotes the local Boltzmann weight for the plaquette spins $\{ S_i \}$ ($i=1\sim4$);
\begin{equation}
\label{local_Boltzmann_weight}
W_{S_1 S_2}^{S_3 S_4} =
\exp \left( - \left(
   -\frac{J_{NN}}{4}(S_1 S_2 +S_2 S_4+S_4 S_3+S_3 S_1)
 - \frac{J_{NNN}}{2}(S_1 S_4 + S_2 S_3)
 - \frac{J_\square}{2}S_1 S_2 S_3 S_4   \right)  \right)    
   .
\end{equation}
Notably enough,
the component $T^{(leg)}$ is nothing but a transfer matrix for the $d=2$ Ising model.
The remaining components $T^{(planar)}$ and $T^{(rung)}$ 
introduce the $\sqrt{N}$th-neighbor couplings,
and raise the dimensionality effectively to $d=3$.

The component $T^{(planar)}$ is given by,
\begin{equation}
\label{T_planar}
T_{ij}^{(planar)} (v) = \langle i| A P^v |i \rangle  ,
\end{equation}
with
\begin{equation}
v=\sqrt{N},
\end{equation}
where the matrix $P$ denotes the translation operator.
Namely, the state $P |i\rangle$ represents a shifted configuration $\{ S(i,m+1) \}$.
Hence,
the insertion of $P^{\sqrt{N}}$ introduces the $\sqrt{N}$th-neighbor interactions
among the $N$ spins \cite{Novotny90}.
Similarly, we propose the following expression for the component $T^{(rung)}$;
\begin{equation}
\label{T_rung}
T^{(rung)}_{ij} (v) =
  \left( \langle i| \otimes \langle j| \right) B 
      \left( \left(P^v |i\rangle\right) 
   \otimes \left(P^v |j\rangle\right) \right)  ,
\end{equation}
with,
\begin{equation}
 \left(  \langle i| \otimes \langle j| \right) B 
    \left(  |k\rangle \otimes |l\rangle \right) = 
  \prod_{m=1}^{N} W_{S(i,m) S(j,m)}^{S(k,m) S(l,m)} .
\end{equation}
The meaning of the formula would be apparent from Fig. \ref{figure8}.

The above formulations are already reported in Ref. \cite{Nishiyama04}.
In the following, we propose a number of additional improvements.
First, we symmetrize the transfer matrix with the replacement
\cite{Novotny92},
\begin{equation}
\label{modification1}
T(v) \to T(v) \odot T(-v)   .
\end{equation}
Correspondingly, we substitute the strength of the coupling
constants 
$J_\alpha \to J_\alpha /2$ in order to compensate the above duplication.
Apparently, with the symmetrization, the symmetry of
descending ($m=N,N-1,\dots $) and ascending ($m=1,2,\dots$) directions
is restored.
Moreover,
we implement the following symmetrizations,
\begin{equation}
\langle i | A P^v | i \rangle \to 
 \langle i| A P^v |i \rangle  \langle i | P^{-v} A | i \rangle    ,
\end{equation}
and,
\begin{eqnarray}
\label{modification3}
  \left( \langle i| \otimes \langle j| \right) B 
      \left( \left(P^{v} |i\rangle\right) 
   \otimes \left(P^{v} |j\rangle\right) \right)   \to   \nonumber \\
  \left( \langle i| \otimes \langle j| \right) B 
      \left( \left(P^{v} |i\rangle\right) 
   \otimes \left(P^{v} |j\rangle\right) \right)  
   \left( (\langle i| P^{-v})\otimes (\langle j| P^{-v})\right)  
  B    \left( |i\rangle 
   \otimes |j\rangle \right)  ,
\end{eqnarray}
as to Eqs. (\ref{T_planar}) and (\ref{T_rung}), respectively.
Again, we have to redefine the coupling constants to compensate the duplication.


\end{document}